\documentclass[aps,prb,reprint,groupedaddress]{revtex4-1}

\usepackage{graphicx}
\usepackage{amsmath} 
\usepackage{color}

\begin{document}

\title{Resonant electron tunneling in a tip-controlled potential landscape}

\author{Nikola Pascher$^{1}$}
\email[]{npascher@phys.ethz.ch}
\author{Flavia Timpu$^{1}$}
\author{Clemens R\"ossler$^{1}$}
\author{Thomas Ihn$^{1}$}
\author{Klaus Ensslin$^{1}$}
\author{Christian Reichl$^{1}$}
\author{Werner Wegscheider$^{1}$}

\affiliation{Solid State Physics Laboratory, ETH Zurich, 8093 Zurich, Switzerland}


\begin{abstract}
By placing the biased tip of an atomic force microscope at a specific position above a semiconductor surface we can locally shape the potential landscape. Inducing a local repulsive potential in a two dimensional electron gas near a quantum point contact one obtains a potential minimum which exhibits a remarkable behavior in transport experiments at high magnetic fields and low temperatures. In such an experiment we observe distinct and reproducible oscillations in the measured conductance as a function of magnetic field, voltages and tip position. They follow a systematic behavior consistent with a resonant tunneling mechanism. From the periodicity in magnetic field we can find the characteristic width of this minimum to be of the order of 100 nm. Surprisingly, this value remains almost the same for different values of the bulk filling factors, although the tip position has to be adjusted by distances of the order of one micron.
\end{abstract}

\pacs{}

\maketitle

\newpage
\section{Introduction}
Charging effects in the integer and fractional quantum Hall regime prove to be powerful tools to probe the properties of the respective particles. These effects offer for example the possibility to measure the charge of the particles \cite{Kou2012,McClure2012,Willett2009}. Furthermore they are discussed as an option to prove possible non-abelian statistics of the 5/2-quasiparticle with a charge of 1/4 \cite{Nayak2008} which would make them a strong candidate to realize topological quantum computation. Fundamental aspects make the underlying physics very rich. It has been shown that many other properties can be efficiently explored by operating electron interferometers and quantum dots in the quantum Hall regime, e.g. to demonstrate the importance of interactions \cite{Rosenow2007,Ihnatsenka2009a,Ofek2010} or to estimate the velocity and dephasing rate inside a quantum Hall edge channel \cite{McClure2009}. Charging effects manifest themselves in an oscillatory pattern in the dependence of the conductance on both the magnetic field and a gate-voltage which influences the size of the confined geometry \cite{Kou2012,Ofek2010,Halperin2011,Zhang2009}. Oscillations in rather small structures are usually dominated by a Coulomb charging- or resonant tunneling mechanism, while large interferometers can be understood based on an Aharanov-Bohm mechanism \cite{Kataoka1998,Zhang2009,Halperin2011}.\\

Several different options exist to confine the current carrying particles in such a way that the above mentioned phenomena can be observed. The crucial point in all these experiments is an isolated island of compressible electron liquid, created in a two dimensional electron gas (2DEG) in a magnetic field. This can be done either by forming a quantum dot or an anti-dot \cite{Zhang2009,Kou2012}. The experimental observations are similar in both cases. All experiments reported up to now used lithographic gates to confine the electron paths into certain geometries. The parameter space is thus spanned basically by the voltages which are applied to the gates, the source-drain bias voltage and the magnetic field. Previous scanning probe experiments used predefined structures and tried to explore the underlying physics with real space resolution \cite{Martins2013}. In our experiments we define the potential landscape with the help of a quantum point contact (QPC) in combination with the biased tip of an atomic force microscope, offering an additional parameter directly influencing the geometry and thus obtaining an additional degree of freedom to tune the structure.

\begin{figure}[]
\centering
\includegraphics[]{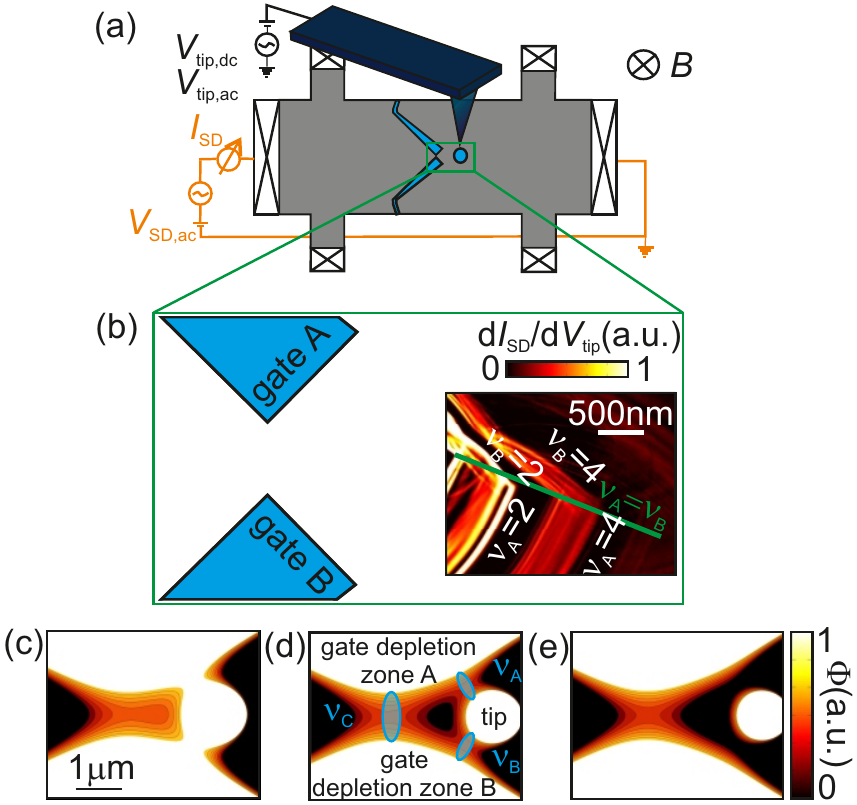}
\caption{(a) Overview over the experimental setup. The source-drain current $I_\mathrm{SD}$ is recorded as a function of tip position. The QPC top gates are shown in blue. The QPC forms at the minimum distance of the two top gates, which is 800 nm. A small ac-modulation is applied to the tip, so that the transconductance $dI_\mathrm{SD}/dV_\mathrm{tip}$ can be measured directly. (b) Typical differential conductance map measured at a bulk filling factor of $\nu_\mathrm{Bulk}=4$, corresponding to an external magnetic field of 2~T. The scan frame is oriented parallel to the QPC gap. The line of high symmetry where $\nu_\mathrm{A}=\nu_\mathrm{B}$ (marked in green) is tilted, likely due to a weak disorder potential. (c)--(e) Model calculation of the potential landscape consisting of a superposition of a saddle point for the QPC and a Lorentzian shape for the tip. (c) If the tip is placed close to the QPC it blocks electron transport. (e) If it is placed far away it has practically no influence. (d) In an intermediate case a potential minimum can form in between the tip and the constriction.}
\label{figure1}
\end{figure}

\section{Sample and experimental setup}
The experimental setup is schematically depicted in Fig. 1(a). We use a GaAs-AlGaAs heterostructure wafer hosting a 2DEG with an electron density of $1.9 \times 10^{11}$~cm$^{-2}$ and a mobility of $3.5 \times 10^6~$cm$^2$/Vs. The 2DEG is placed 120 nm below the surface. Top gates to define a QPC were patterned using electron beam lithography on top of a wet-etched Hall bar structure. We apply an ac-source-drain voltage of 20~$\mu$V symmetrically with respect to ground and measure the source-drain current $I_\mathrm{SD}$ at the same time. The gold-top gates have an opening angle of 90$^{\circ}$ and are spaced by 800 nm at the point of minimum distance (see Fig. 1(a)). We use the conducting tip of a home-built cryogenic scanning force microscope (AFM) operating in a dilution refrigerator at a base temperature of 40 mK to locally perturb electron transport. For this purpose the tip is scanned at a height of 90 nm above the sample surface and a negative dc-voltage $V_\mathrm{tip}=-3.5$ V is applied unless stated otherwise. This leads to a disc-shaped region of total depletion in the electron gas underneath, with a radius of about 1.2 $\mu$m \cite{Pascher2013}. We apply a small ac-voltage with an amplitude of 50~mV to the tip, in order to measure the transconductance $dI_\mathrm{SD}/dV_\mathrm{tip}$. In most of the existing literature charging effects are directly visible as oscillations in $I_\mathrm{SD}$ \cite{Zhang2009,Ofek2010}. In our measurements the oscillations in $I_\mathrm{SD}$ are very subtle and superimposed on a strongly varying background. It is therefore necessary to measure $dI_\mathrm{SD}/dV_\mathrm{tip}$ in order to enhance their visibility. The electron temperature in the experiment is about 170 mK, and a magnetic field is applied perpendicular to the 2DEG. \\

\section{Creating the potential minimum}
Figures 1(c), (d) and (e) show schematically the potential landscape which is induced by the gates and the tip. Figures 1(c) and (e) show the extreme cases, where the tip is so close to the QPC, that it totally blocks transport or the tip is so far away, that it has practically no influence on the recorded current. The most interesting case occurs at intermediate tip-positions shown in Fig. 1(d), where two QPCs A and B form between the tip and the two top gates A and B \cite{Pascher2013}. We can assign a different filling factor to each QPC. We name the filling factor in the center of the lithographic QPC $\nu_\mathrm{C}$ and the filling factors of the QPCs which are caused by the tip $\nu_\mathrm{A}$ and $\nu_\mathrm{B}$. The measured current is

\begin{equation}
	I_\mathrm{SD}=\frac{e^2}{h}\nu V_\mathrm{SD},
\end{equation}

where 

\begin{equation}
	\nu=\mathrm{min}[\nu_\mathrm{C},\mathrm{max}(\nu_\mathrm{A},\nu_\mathrm{B})].
\end{equation}

A special situation arises, when all three filling factors are the same. As shown in Fig. 1(d) a potential minimum between the gates and the tip forms, which leads to localized states. The local filling factor $\nu_{loc}$ inside the potential minimum can be tuned in a parameter space spanned by the external magnetic field $B$, the gate-voltage $V_\mathrm{gates}$, the tip voltage $V_\mathrm{tip}$, and the tip position.\\

\begin{figure*}
\centering
\includegraphics[]{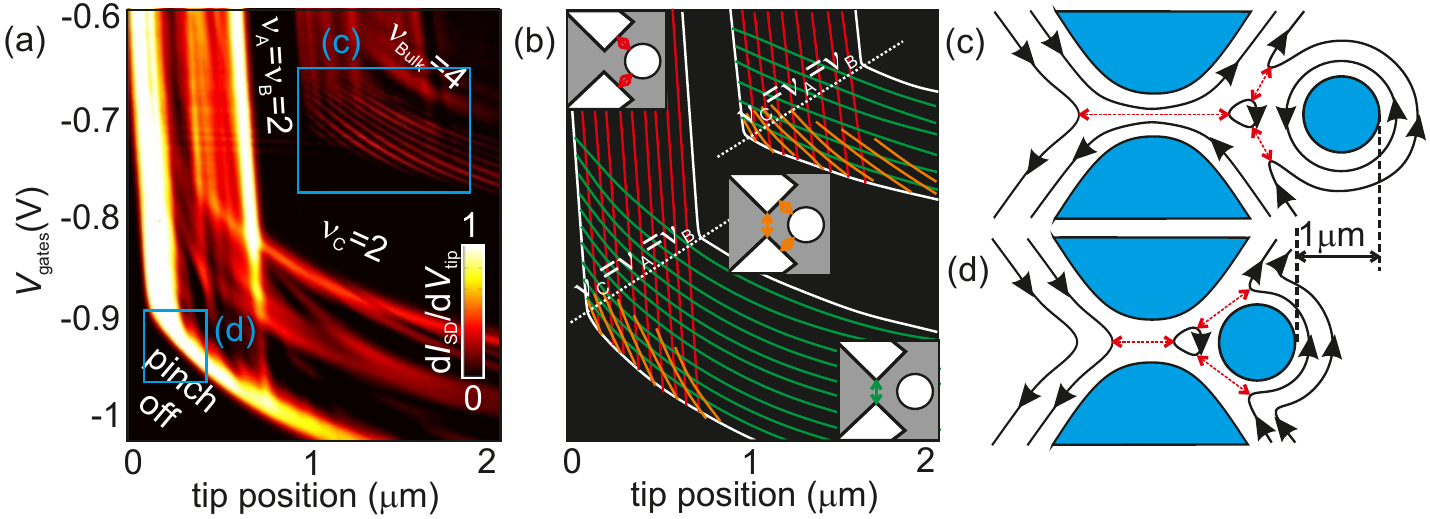}
\caption{(a) The green line, where $\nu_\mathrm{A}=\nu_\mathrm{B}$ in Fig. 1(b) is repeatedly scanned, while the gates-voltage $V_\mathrm{gates}$ is changed after each line is finished. The resulting $dI_\mathrm{SD}/dV_\mathrm{tip}$ false-color plot shows two prominent L-shaped features, decorated with interference patterns at the positions of the blue frames. Depending on gate voltage and tip position the conductance changes from total pinch-off to the bulk filling factor of 4. (b) Schematic plot to explain the observations in (a). Features which are parallel to the red lines are mainly caused by the QPCs which form in between the tip and the QPC gates. The green lines mark the regime, where the QPC in between the lithographic gates starts to dominate. Features which cut the so-constructed coordinate systems (shown in orange) are caused by an island which is confined by all three QPCs. (c) Describes the situation for the upper blue in (a). One spin-degenerate edge channel is completely transmitted, one is reflected. A closed loop edge states forms. In (d), describing the lower blue frame in Fig. 2(a), two spin degenerate edge channels are reflected. A confined loop with the a similar size forms in the potential minimum. The tip has to be laterally moved by about 1~$\mu$m to change from one situation to the other.}
\label{figure2}
\end{figure*}

\section{Finding the right spot in parameter space}
\subsection{The symmetry line}
The correct position for the tip has to be found, where $\nu_\mathrm{A}=\nu_\mathrm{B}=\nu_\mathrm{C}$ and the puddle will form. In a first step we aim at finding the line where $\nu_\mathrm{A}=\nu_\mathrm{B}$. For that purpose a constant voltage is applied to the QPC-gates in order to define the QPC. A voltage of $-0.3$ V was chosen, which is enough to deplete the 2DEG underneath \cite{Rossler2011} but keeps $\nu_\mathrm{C}$ maximal. Figure 1(b) shows an area-scan which was taken at a magnetic field of 2~T (bulk filling factor $\nu_\mathrm{Bulk}=4$) at a position about 2~$\mu m$ away from the QPC-gap with the $y$-axis of the scan frame oriented parallel to the QPC gap. The colors encode the transconductance $dI_\mathrm{SD}/dV_\mathrm{tip}$ measured with respect to the tip-voltage. To clarify the experimental geometry, the scanning gate microscopy (SGM)-image is put into relation to the position of the lithographic gates in real space, as indicated by the green frame. In this constellation, this means that $\nu_\mathrm{C}$ is maximal and according to Eq. (2) $\nu_\mathrm{A}$ and $\nu_\mathrm{B}$ dominate. \\

The current $I_\mathrm{SD}$ saturates at the bulk-value which is fixed to the bulk filling factor $\nu_\mathrm{Bulk}=4$. In the $dI_\mathrm{SD}/dV_\mathrm{tip}$-map shown in Fig. 1(b) plateaus in the conductance as a function of tip position show up as black stripes. Broad black stripes correspond to perfect transmission of even integer quantum Hall edge states according to Eq. (2). Thus we can assign specific filling factors to the black stripes as indicated in Fig. 1(b). The green line marks a line of symmetry where $\nu_\mathrm{A}=\nu_\mathrm{B}$. This line is not perpendicular to the QPC-gap, either because the left and the right topgate have a slightly different lever arm, or because the local potential landscape is not perfectly symmetric due to irregularities in the lithographic top gates or local effects of impurities \cite{Rossler2011}.\\

\begin{figure*}
\centering
\includegraphics[]{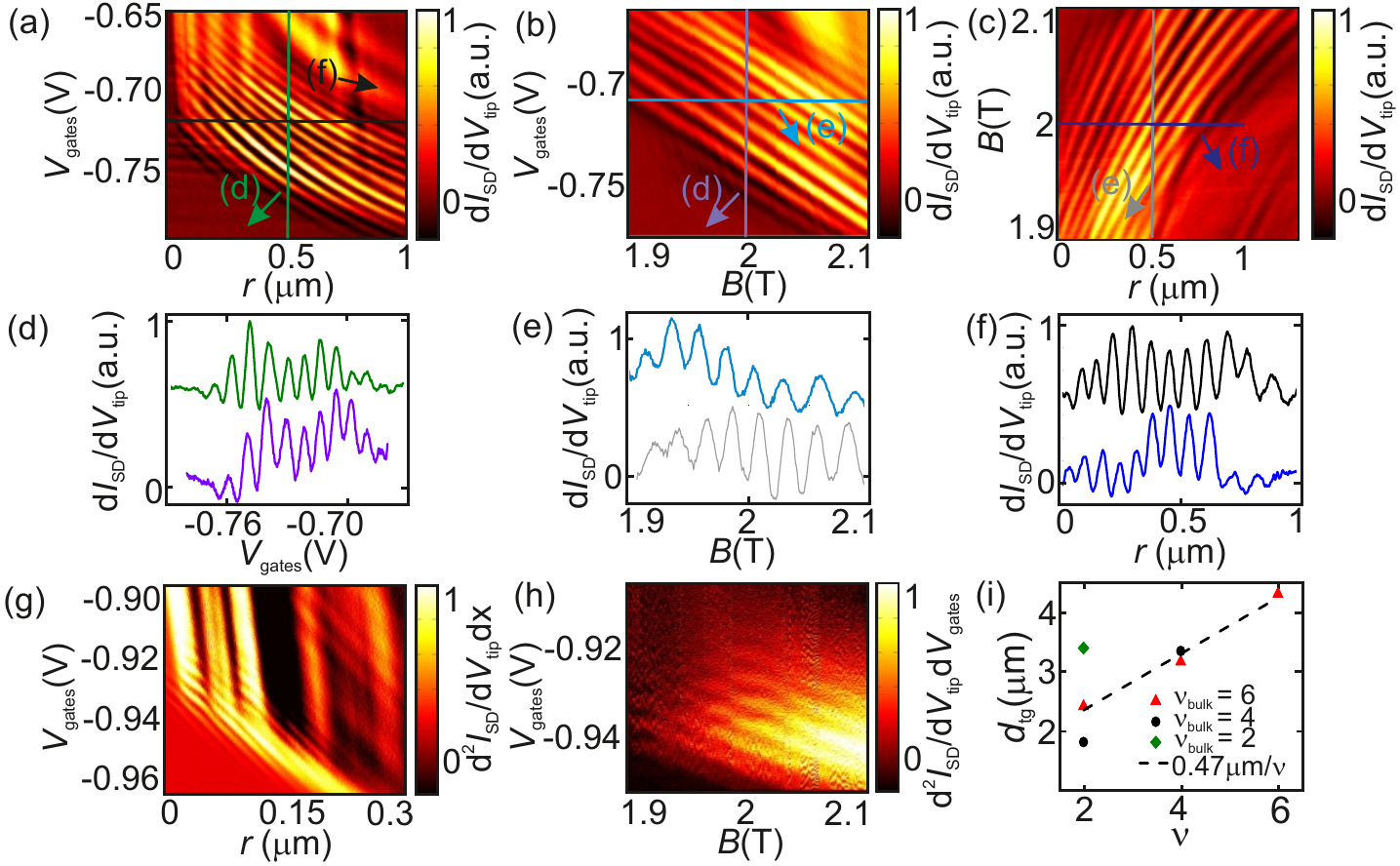}
\caption{(a), (g) Zoom at the position of the blue frame (c) and (d) number in Fig. 2(a). (b), (h) The tip is placed at a fixed position and $dI_\mathrm{SD}/dV_\mathrm{tip}$ is recorded both as a function of $V_\mathrm{gates}$ and $B$. A positively sloped stripe-pattern is resolved. (c) Oscillation pattern as a function of tip position $r$ along the same line as in (a)and $B$. (d) - (f) Linecuts at the colored lines in (a) - (c). (i) Tip position for different filling factors $\nu$. To transmit an additional spin-degenerate edge channel the tip has to be moved by about 1~$\mu$m.}
\label{figure3}
\end{figure*}

\subsection{Closing the quantum point contact}
Now that the line where $\nu_\mathrm{A}=\nu_\mathrm{B}$ is found, we need to find the tip position where $\nu_\mathrm{A}=\nu_\mathrm{B}=\nu_\mathrm{C}$. The line is repeatedly scanned, while the voltage on the QPC-gates is changed after each line is finished. Exemplary results are presented in Fig. 2 for a magnetic field of 2~T corresponding to a bulk filling factor $\nu_\mathrm{Bulk}=4$ (Fig. 2(a)) together with schematics to explain the prominent features (Figs. 2(b) - (d)). Different bulk filling factors show qualitatively similar results.\\

In these line scan-images we can distinguish two different regimes, as illustrated in Fig. 2(b): At relatively low gates-voltages and small distance between QPC and tip (top left corner of Fig. 2(b)) $\nu_\mathrm{A}$ and $\nu_\mathrm{B}$ dominate, because $\nu_\mathrm{A}=\nu_\mathrm{B}<\nu_\mathrm{C}$. In this regime, the pattern does not depend on the gate-voltage, yet the tip position has a huge influence. This regime is marked by the red parallel lines in Fig. 2(b). At larger gate voltages and larger tip-QPC separation (bottom right corner of Fig. 2(b)) $\nu_\mathrm{C}$ dominates, because $\nu_\mathrm{A}=\nu_\mathrm{B} > \nu_\mathrm{C}$, as illustrated by the green lines in Fig. 2(b). In this regime the dependence on the tip position is less pronounced than before. The shape of this curve as a function of position and voltage reflects the tip-induced potential.\\

In this parameter space we find lines where $\nu_\mathrm{A}=\nu_\mathrm{B}=\nu_\mathrm{C}$ which form the boundaries between two cases (see Fig. 2(b)). This is where the above mentioned potential minimum between tip and QPC can be found. We find resonances at every transition between local filling factors (see blue frames in Fig. 2(a)). The transition between $\nu=2$ and $\nu=4$ is shown in Fig. 2(c): One spin degenerate edge channel is fully transmitted, one is reflected by the SGM-tip. At the transition between filling factors $\nu=0$ and $\nu=2$ the situation (see Fig. 2(d)) is similar, although the oscillation pattern is not clearly visible in Fig. 2(a).

\section{Parameter dependencies of oscillations}
In order to characterize the parameter-dependencies of the observed oscillations quantitatively, we zoom into the regions within the blue frames in Fig. 2(a) at the bulk filling factor of 4. These correspond to the transition regions between filling factors 2 and 4 in the case of frame (c) and between 0 and 2 for the case of frame (d). In both cases the zooms reveal an oscillatory pattern, as shown in Figs. 3(a) and (g). At the transition between pinch off and the local filling factor 2 the signal is rather weak, which makes it necessary to plot the second derivative in order to achieve significant contrast. 

\begin{figure}
\centering
\includegraphics[]{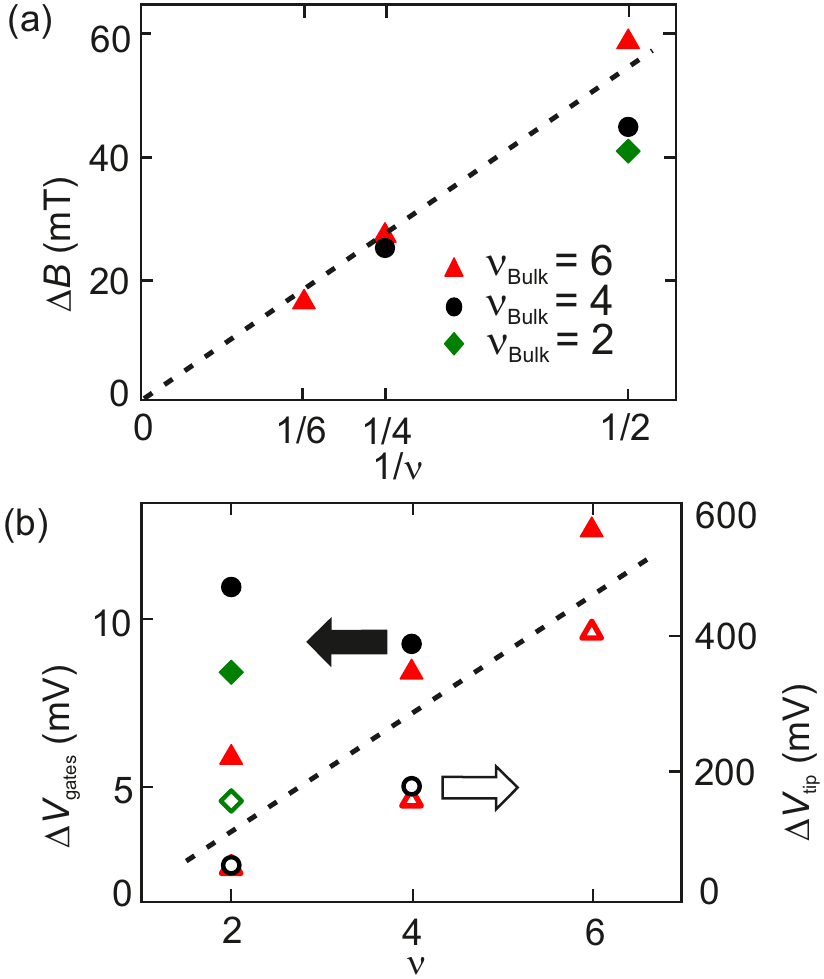}
\caption{(a) The magnetic field period $\Delta B$ scales with $1/\nu_\mathrm{QPC}$. (b) The periods $\Delta V_\mathrm{gates}$ and $\Delta V_\mathrm{tip}$ increase with $\nu_\mathrm{QPC}$. The tip or the gates lead to a change in the local density, energy structure and the shape of the potential dip. Full symbols mark $\Delta V_\mathrm{gates}$ (left hand scale, empty symbols mark $\Delta V_\mathrm{tip}$, right hand side scale)}
\label{figure4}
\end{figure}

\subsection{Magnetic field dependence of oscillations}
When the tip is placed at a fixed position, the dependence of the oscillations on $B$, $V_\mathrm{gates}$ and $V_\mathrm{tip}$ can be measured. The resulting oscillation periods are named $\Delta B$, $\Delta V_\mathrm{gates}$ and $\Delta V_\mathrm{tip}$. The distance between tip and QPC-gap on the high symmetry line $\nu_\mathrm{A}=\nu_\mathrm{B}$ is $d_\mathrm{tg}$ and needs to be changed to adjust different filling factors. In Fig. 3(b) the dependence of the oscillatory pattern as a function of $B$ and $V_\mathrm{gates}$ is plotted. A stripe-pattern with a positive slope is observed. The periods $\Delta B$ and $\Delta V_\mathrm{gates}$ can be read from the linecuts, as shown in Figs. 3(d) - (f). Sweeping the voltage on the tip instead of the gates leads to qualitatively similar pictures (not shown). The same measurements (Figs. 3 (g), (h)) were performed at the position of the blue frame (d) in Fig. 2(a). \\

To illustrate the reproducibility of the measurement results, traces which were taken in different measurements as a function of the same parameter are compared in Figs. 3(d) - (f). The curves were vertically offset for clarity. Lines in Figs. 3(a) - (c) encode which traces are shown in (d) - (f). Matching colors were chosen. The positions of the peaks agree very well, their intensities are slightly different. These measurements were taken in the order of one day apart, which leads to these small changes in the sample which give rise to small differences.

\begin{table*}[]
\caption{Summary of the obtained values for different bulk filling factors $\nu_\mathrm{Bulk}$, different local filling factors inside the QPC $\nu_\mathrm{QPC}$ and different tip-QPC-distances $d_\mathrm{tg}$. The oscillation period as a function of magnetic field $\Delta B$ allows to calculate the radius of the island $R$ according to Eq. (3). Periods can be observed in $\Delta V_\mathrm{gates}$ and $\Delta V_\mathrm{tip}$. The errors are estimated according to the standard deviation for period values which are extracted from traces like shown in Fig. 3(d) - (f). The error in $d_\mathrm{tg}$ comes from an uncertainty in deducing the exact position of the QPC-gap during SGM-measurements.}
\begin{tabular}{ccccccc} \hline \hline
			$\nu_\mathrm{Bulk}$&$\nu$&$\Delta B$(mT)&$\Delta V_\mathrm{gates}$(mV)&$\Delta V_\mathrm{tip}(mV)$&$d_\mathrm{tg}$($\mu m$)&$R$(nm)\\ 
			 & &$\pm$2 mT&$\pm$1~mV&$\pm$10~mV&$\pm$0.02&$\pm$3~nm\\ \hline
			6 & 6 ... 4 & 16 & 13 & 403 & 4.32 & 117\\ 
			  & 4 ... 2 & 27 & 8 & 150 & 3.18 & 110\\
			  & 2 ... 0 & 59 & 5 & 48 & 2.43 & 106\\ \hline
			4 & 4 ... 2 & 25 & 9 & 171 & 3.36 & 115\\ 
			  & 2 ... 0 & 45 & 11 & 52 & 1.83 & 121\\ \hline
			2 & 2 ... 0 & 41 & 8 & 149 & 3.41 & 127\\ \hline \hline
\end{tabular}
\end{table*}

\subsection{Interpretation}
In the existing literature several possible mechanisms are discussed to understand conductance oscillations in confined geometries in quantum Hall systems. The positive slope in the conductance maps as a function of the magnetic field and gate voltage observed in the present investigation shows, that an increase of the magnetic field needs to be compensated by an increased gate voltage, as it would be the case for a Coulomb charging mechanism \cite{Rosenow2007,Zhang2009}. Fundamental to our analysis will be the assumption that within each oscillation period, one electron is added to the island. \\

At the position of the potential minimum there is one edge channel which circulates in a closed loop, as illustrated in Figs. 2(c) and (d). A change in magnetic field changes the energy of the Landau levels and their degeneracy. A change in magnetic field will change the number of electrons on the localized island, which means that

\begin{equation}
	\Delta B = \frac{1}{\nu_\mathrm{loc}}\frac{\Phi_0}{A},
\end{equation}

where A is the area of the island and $\nu_\mathrm{loc}$ is its filling factor. A change of the magnetic field $\Delta B$ by one flux quantum $\Phi_0$ per area $A$ leads to $\nu_\mathrm{loc}$ additional electrons on the island. Whenever two consecutive occupation numbers are degenerate, Coulomb blockade of the island is lifted which gives rise to conductance oscillations with a period $\Delta B$. Changing $V_\mathrm{tip}$ or $V_\mathrm{gates}$ leads to a similar effect on the number of electrons on the island. Electrons will tunnel to the transmitted edge channels and conductance oscillations will be observed. \\

Figure 3(c) shows the evolution of the oscillation pattern as a function of tip position along the line $\nu_\mathrm{A}=\nu_\mathrm{B}$. Moving the tip has an effect on the area of the localized region, if the tip is moved further away from the QPC, the area should increase. Thus the first effect which influences the shape of the oscillation pattern is the non-linear tail of the Lorentzian potential, which is induced by the tip. The second effect describes the dependence on a change of the magnetic field \cite{Halperin2011,Baer2013}. If $B$ is increased, the population of the highest Landau level is decreased. This reduced electron density in the edge channels leads to an increased oscillation period at higher magnetic fields. \\

All oscillation periods which were extracted are summarized in Tab. 1. Similar measurements as presented in Fig. 3 for $\nu_\mathrm{Bulk}=4$ were also performed for $\nu_\mathrm{Bulk}=6$ and $\nu_\mathrm{Bulk}=2$ (not shown). The quoted errors are the standard deviations for period values which are extracted from traces like those shown in Fig. 3(d) -- (f). A small systematic error is included, which is due to a linear increase of the peak spacing with higher gate- or tip-voltages. The error in $d_\mathrm{tg}$ describes the uncertainty in deducing the exact position of the QPC-gap during SGM-measurements. The periods $\Delta B$ as a function of 1/$\nu_\mathrm{QPC}$ show a linear behavior (see Fig. 4 (a)), consistent with Eq. (3). With the help of Eq. (3) we can estimate the size $R$ of the island from $\Delta B$ which is found to be always very similar (see Tab. 1). Assuming an approximately circular shape, we can calculate a radius $R$=(120$\pm$10) nm. With the help of a capacitor model we can estimate the charging energy of the island with this radius to be $E_c=1.6\pm0.14$~meV. As this energy is needed to change the charge on the island by one electron, the axes $V_\mathrm{gates}$, $V_\mathrm{tip}$ and $B$ can be directly translated into energies via the peak spacings. From the full width at half maximum of the resonance peaks we can estimate the strength of the tunneling coupling $\Gamma$. A Lorentzian fit to the peaks yields a value $\Gamma=(1.7\pm0.4$)~meV. Thus the system is in a regime, where the charge on the island is still approximately quantized, but the island is strongly coupled to the transmitting edge channels. No clear Coulomb blockade diamonds can be measured in this regime, as we verified experimentally. \\

In Fig. 4(b) we show the oscillation periods $\Delta V_\mathrm{gates}$ and $\Delta V_\mathrm{tip}$ as a function of filling factor. If the only effect of the tip or the gates is to change the electron number on the island, the periods $\Delta V_\mathrm{gates}$ and $\Delta V_\mathrm{tip}$ should be constant for all filling factors \cite{Zhang2009}. In our case changing $V_\mathrm{gates}$, $V_\mathrm{tip}$ or $d_\mathrm{tg}$ does not only change the electron number, but also the size of the island. Thus the periods $V_\mathrm{gates}$ and $V_\mathrm{tip}$ vary.\\

To adjust a particular filling factor $\nu= \nu_\mathrm{A} = \nu_\mathrm{B} = \nu_\mathrm{C}$ at a given bulk filling factor $\nu_\mathrm{Bulk}$, the tip needs to be moved to a particular position $d_\mathrm{tg}$. These positions as a function of filling factor are plotted in Fig. 3(i). The points can be fitted with a line showing, that the tip has to be moved by about 1 $\mu$m to enable perfect transmission of an additional spin degenerate edge channel. This corresponds to a distance of 700 nm between the tip and the right or left topgate. This situation is schematically illustrated in Figs. 2(c) and (d). This allows us to estimate the width of one spin degenerate edge channel in the local potential landscape to be about 350 nm.\\

\section{Conclusions}
Complementing previous experiments \cite{Kou2012,Ofek2010,Halperin2011,Zhang2009} our results demonstrate that an electrostatic potential minimum can be established and tuned using the metallic tip of a scanning force microscope. This gives direct access to the spatial degrees of freedom from the experimental point of view. Distinct and reproducible oscillations in the measured conductance as a function of magnetic field, voltages and tip position follow a systematic behavior consistent with a resonant tunneling mechanism. The periodicity in magnetic field allows to estimate the characteristic width of this minimum to be of the order of 100 nm. This value remains constant for different filling factors, even though the tip has to be moved by the surprisingly large distance of 1 $\mu$m, which allows us to estimate the width of one spin degenerate edge channel to be about 350 nm.

\section*{Acknowledgments}
The authors acknowledge the Swiss National Science Foundation, which supported this research through the National Centre of Competence in Research "Quantum Science and
Technology" and the Marie Curie Initial Training Action (ITN) Q-NET 264034. We thank B. Rosenow and C. Marcus for fruitful discussions.

\end{document}